\documentclass[twocolumn]{autart}    

\usepackage[english]{babel}
\usepackage[T1]{fontenc}
\usepackage[dvips]{graphicx} 
\usepackage[dvips]{epsfig} 
\usepackage[cmex10]{amsmath} 
\usepackage{amssymb}  
\usepackage[tight, footnotesize]{subfigure}
\usepackage{url}
\usepackage{bm}
\usepackage{subfigure}
\usepackage{theorem}

\usepackage{color}

\usepackage[retainorgcmds]{IEEEtrantools}
\usepackage[noadjust]{cite}

\newtheorem{hypothesis}{Assumption}

\newtheorem{remark}{Remark}

\theoremstyle{definition}

\theoremstyle{definition}
\newtheorem{proposition}{Proposition}

\newenvironment{system}[1][rCL]
{\left\lbrace\begin{IEEEeqnarraybox}[][c]{#1}}
{\end{IEEEeqnarraybox}\right.}

\newcommand{\rot}{\mathrm{rot}}

\graphicspath{figures/}

\begin{document}

\begin{frontmatter}

\title{Modeling for Control of Symmetric Aerial Vehicles\\ Subjected to Aerodynamic Forces}

\author[INRIA]{Daniele Pucci}
\author[I3S]{Tarek Hamel}
\author[UPMC]{Pascal Morin}
\author[INRIA]{Claude Samson}
\address[INRIA]{firstname.surname@inria.fr, INRIA Sophia Antipolis M\'{e}diterran\'{e}e, 2004 Route des Lucioles BP 93,
Sophia Antipolis, France}  
\address[I3S]{thamel@i3s.unice.fr, I3S Universit\'e Nice Sophia-Antipolis, 2000 Route des Lucioles BP 121, 06902 Sophia-Antipolis, France}             
\address[UPMC]{morin@isir.upmc.fr, ISIR, Universit\'e Pierre et Marie Curie (UPMC), 75005 Paris, France  }        

\graphicspath{{figures/}}

\begin{keyword}
Aerial Vehicles, Aerodynamic Forces, Modeling, Nonlinear Control.
\end{keyword}

\begin{abstract}
This paper participates in the development of a unified approach to the control of aerial vehicles with extended flight envelopes. More precisely,
modeling for control purposes of a class of thrust-propelled aerial vehicles subjected to lift and drag aerodynamic forces is 
addressed assuming a rotational symmetry of the vehicle's shape about 
the thrust force 
axis.
A condition upon aerodynamic characteristics that allows one to recast the control
problem into the simpler case of a spherical vehicle is pointed out. Beside showing how to adapt nonlinear controllers developed for this latter case, the paper extends a previous work by the authors in two directions.
First, the 3D case is addressed whereas only motions in a single vertical plane was considered. Secondly, the family of models of aerodynamic forces for which the aforementioned transformation holds is enlarged.
\end{abstract}

\end{frontmatter}

\section{Introduction}
Feedback
control of aerial vehicles in order to achieve some degree of autonomy remains an active
research domain after decades of studies on the subject. The complexity of aerodynamic effects and the diversity of
flying vehicles partly account for this continued interest. Lately, the  emergence of small vehicles for robotic applications
(helicopters, quad-rotors, etc) has also renewed the interest of the control community for these systems.
Most aerial vehicles belong either to the class of fixed-wing vehicles,
or to that of rotary-wing vehicles.
The first class is mainly composed of airplanes. In this case, weight is compensated for by lift forces
acting essentially on the wings, and propulsion is used to counteract drag forces associated with large
air velocities. The second class contains several types of systems, like helicopters,
ducted fans, quad-rotors, etc. In this case, lift forces are usually not preponderant and the \emph{thrust force}, produced by one or several propellers, has also to compensate for the vehicle's weight. These vehicles are usually referred to as Vertical Take-Off
and Landing vehicles (VTOLs) because they can perform stationary flight (hovering). On the other hand, energy consumption
is high due to small lift-to-drag ratios. By contrast, airplanes cannot (usually) perform stationary flight,
but they are much more efficient energetically than VTOLs in cruising mode.

Control design techniques for airplanes and VTOLs have developed along different directions and suffer from specific limitations. Feedback control of airplanes explicitly takes into account lift forces
via linearized models at low angles of attack. Based on these models, stabilization is usually achieved through
linear control techniques \cite{2004_STENGEL}. As a consequence, the obtained stability
is local and difficult to quantify. Linear techniques are
used for hovering VTOLs too, but several nonlinear feedback methods have also been proposed in the last decade to
enlarge the provable domain of stability ~\cite{2005_MCMFM} \cite{1992_HAUSER}~\cite{2003_RAG}~\cite{2002_MARCONI}. These methods, however, are based on simplified dynamic models that neglect aerodynamic forces. For this reason, they are not best suited to the control of aerial vehicles moving fast or subjected to strong wind variations.
Another drawback of the independent development of control methods for airplanes and VTOLs is the lack of tools
for flying vehicles that belong to both classes. These vehicles are usually referred to as \emph{convertible} because they can perform stationary flight and also benefit from lift
properties at high airspeed via optimized aerodynamic profiles. The renewed interest in such vehicles and their control reflects in the growing number of studies devoted to them
in recent years ~\cite{2007_BENOSMAN}~\cite{2007_FRANK} \cite{2011_NALDI}~\cite{1999_OISHI}, even though the literature in this domain is not much developed yet. One of the
motivations for elaborating more versatile control solutions is that the automatic monitoring of the delicate transitions between stationary flight and cruising modes, in relation to the strong variations of drag and lift forces during these transitions, remains a challenge to these days.
A first step in this direction consists in taking into account drag forces that do not depend on the vehicle's orientation \cite{2009_HUA}, as in the case of spherical bodies.

The present paper essentially aims at extending \cite{2009_HUA} by taking lift forces into account and extending to the 3D case a previous contribution \cite{2011_pucci} concerning vehicles moving in the vertical plane (2D case) which shows how, for a particular class of models of lift and drag aerodynamic forces acting on a wing, it is possible to bring the control problem back to the simpler one of controlling a spherical body subjected to an orientation-independent drag component solely. One can then apply the nonlinear control schemes proposed in \cite{2009_HUA} for which quasi global stability and convergence results are established. The results here reported  thus constitute a contribution to setting the principles of a general nonlinear control framework that applies to many aerial vehicles evolving in a large range of operational and environmental conditions.

The paper is organized as follows. In the background Section \ref{sec:background}, after specifying the notation used in the paper and the actuation framework considered to control the vehicle, general dynamics equations are recalled and motivations for the present study are discussed in relation to the limitations of previous results addressing the case of spherical bodies.
Original results concerning the modeling of aerodynamic forces acting on symmetric bodies and the characterization of a family of models that allow one to recast the control problem into the simpler case of a vehicle subjected to only an orientation-independent drag force are reported in Section~\ref{sec:aerodyForce}. Members of this family are singled out and tuned by using experimental data borrowed from \cite{1965_WAYNE}  and \cite{1971_SAFFEL} for elliptic-shaped and missile-like bodies.
To illustrate the usefulness of these results at the control design level with an example, Section \ref{sec:controlDesign} gives the adapted version of a velocity control scheme proposed in \cite{2009_HUA}. The concluding Section \ref{conclusion} offers complementary remarks and points out research perspectives.

\section{Background}
\label{sec:background}

\subsection{Notation}
\label{sec:notation}
\begin{itemize}
 \item The $i_{th}$ component of a vector $x$ is denoted as $x_i$.
 \item For the sake of conciseness, $(x_1 \vec{\imath} + x_2 \vec{\jmath} +x_3 \vec{k})$ is written as $(\vec{\imath},\vec{\jmath},\vec{k})x$.
 \item $S(\cdot)$ is the skew-symmetric matrix-valued operator associated with the cross product in $\mathbb{R}^3$, i.e. such that $S(x)y=x \times y$, $\forall (x,y) \in \mathbb{R}^3 \times \mathbb{R}^3$.
 \item $\{e_1,e_2,e_3\}$ is the canonical basis in $\mathbb{R}^3$.
 \item \emph{m} is the mass of the vehicle, assumed to be constant, and $G_m$ is the body's center of mass.
 \item $\mathcal{I} = \{O;\vec{\imath}_0,\vec{\jmath}_0,\vec{k}_0\}$ is a fixed inertial frame with respect to (w.r.t.) which the vehicle's absolute pose is measured,
and $\mathcal{B} = \{G;\vec{\imath},\vec{\jmath},\vec{k}\}$ is a frame attached to the body. Observe that $G$ and $G_m$ may not coincide.


\item The body's linear velocity is denoted by $\vec{v} =\frac{d}{dt}\vec{OG}_m= (\vec{\imath}_0,\vec{\jmath}_0,\vec{k}_0)\dot{x}= (\vec{\imath},\vec{\jmath},\vec{k})v$.
\item The linear acceleration vector is $\vec{a}=\frac{d}{dt}\vec{v}$.
\item The body's angular velocity is $\vec{\omega}=(\vec{\imath},\vec{\jmath},\vec{k})\omega$.
\item The vehicle's orientation w.r.t. the inertial frame is represented by the rotation matrix $R$. The column vectors of $R$ are the vectors of coordinates of
$\vec{\imath},\vec{\jmath},\vec{k}$ expressed in the basis of  $\mathcal{I}$.
\item The wind's velocity vector $\vec{v}_w$ is assumed to be the same at all points in a domain surrounding the vehicle, and its components are defined by $\vec{v}_w = (\vec{\imath}, \vec{\jmath}, \vec{k})v_w$. The \emph{airvelocity} $\vec{v}_a  {=} (\vec{\imath},\vec{\jmath},\vec{k})v_a{=}(\vec{\imath}_0,\vec{\jmath}_0,\vec{k}_0)\dot{x}_a$ is defined as the difference between $\vec{v}$ and $\vec{v}_w$. Thus, $v_a = v - v_w$.
\end{itemize}

\subsection{Vehicle's actuation}
\label{sec:actuation}
To cover a large number of actuation possibilities associated with {\em underactuated} aerial vehicles, and work out general principles applicable to many of
them, one must get free of actuation specificities and concentrate on operational common denominators. This leads us to assume, as in \cite{2009_HUA}, that the vehicle's means of actuation
consist of a thrust force $\vec{T}$ along a body-fixed direction, and a torque $\vec{\Gamma}_G$ which allows one to modify the body's instantaneous angular velocity $\vec{\omega}$ at will. In practice, this torque is produced in various ways, typically with secondary propellers (VTOL vehicles), rudders or flaps (airplanes), control moment gyros (spacecrafts), etc. The latter assumption implicitly implies that the torque calculation and the ways of producing this torque can theoretically be decoupled from high-level control objectives. The corresponding requirement is that ``almost'' any desired angular velocity can physically be obtained ``almost'' instantaneously. Under these assumptions, the control of the vehicle relies upon the determination of four input variables, namely the thrust intensity and the three components of $\vec{\omega}$. The following complementary assumption about the thrust force is made.

\begin{hypothesis}
\label{hy:actuation}
The thrust force $\vec{T}$ is parallel to the vector $\vec{k}$, i.e. $\vec{T} = -T\vec{k}$ with $T$ denoting the thrust intensity.
\end{hypothesis}
The minus sign in front of the equality's right-hand side is motivated by a sign convention, also used in \cite{2009_HUA}.

\subsection{Vehicle's dynamics}
\label{subsec:systemmodeling}

The external forces acting on the body are composed of the weight vector $m \vec{g}$ and the sum of aerodynamic
forces denoted by $\vec{F}_a$. In view of Assumption~\ref{hy:actuation}, applying the fundamental theorem of mechanics yields the following
equations of motion:
\begin{IEEEeqnarray}{RCL}
	 m\vec{a} &=&  m\vec{g} +\vec{F}_a -T\vec{k},
	\label{eq:newton} \\
        \frac{d}{dt} (\vec{\imath},\vec{\jmath},\vec{k}) &=& \vec{\omega} \times (\vec{\imath},\vec{\jmath},\vec{k}),
	\label{eq:newtonM}
\end{IEEEeqnarray}
with $T$ and $\omega$ the system's control inputs. The motion equation \eqref{eq:newton} points out the role of the aerodynamic force $\vec{F}_a$ in obtaining the body's linear acceleration vector $\vec{a}$. It shows, for instance, that to move with a constant reference velocity the controlled thrust vector $T \vec{k}$ must be equal to the resultant external force $\vec{F}_{ext} :=m \vec{g}+\vec{F}_a$. 
When $\vec{F}_a$ does not depend on the vehicle's orientation, as in the case of spherical bodies subjected to orientation-independent drag forces only, the resultant external force does not depend on this orientation either. The control strategy then basically consists in aligning the thrust direction $\vec{k}$ with the direction of $\vec{F}_{ext}$ (orientation control with $\omega$) and in opposing the thrust magnitude to the intensity of $\vec{F}_{ext}$ (thrust control with $T$). The almost-globally stabilizing controllers proposed in  \cite{2009_HUA} illustrate this strategy. However, the production of lift and drag forces that depend on the vehicle's orientation may significantly complexify this strategy. In particular, the resultant force $\vec{F}_{ext}$ being now orientation-dependent, the existence and uniqueness of the equilibrium in terms of the vehicle's orientation is no longer systematic, and the stabilization of such an equilibrium, when it is locally unique, can be very sensitive to thrust 
orientation variations. 
As a matter of fact, the capacity of calculating the direction and intensity of $\vec{F}_a$ at every time-instant --already a quite demanding requirement-- is not sufficient to design a control law capable of performing well in (almost) all situations. Knowing how this force changes when the vehicle's orientation varies is needed, but is still not sufficient. An original outcome of the present study is precisely to point out the existence of a generic set of aerodynamic models that allow one
to recast the control problem into the one of controlling a spherical body for which strong stability and convergence results can be demonstrated. Of course, the underlying assumptions are that these models reflect the physical reality sufficiently well and that the corresponding aerodynamic forces can be either measured or estimated on line with sufficient accuracy.

\section{Modeling of aerodynamic forces}
\label{sec:aerodyForce}

\subsection{Models of lift and drag forces}

Working out a functional model of aerodynamic forces from celebrated \emph{Navier$-$Stokes nonlinear partial differential equations} governing the interactions between a solid body and the surrounding fluid is beyond the authors domain of expertise, all the more so that spatial integration of these equations over the shape of an object
does not yield closed-form expressions except in very specific cases. Notwithstanding the delicate and complex issues associated with turbulent flows --a side effect of which is the well known {\em stall} phenomenon--  for which no general complete theory exists to our knowledge. We thus propose to take here a different route by combining a well-accepted general expression of the intensity of aerodynamic forces with geometric considerations based on the body's symmetry properties. To be more precise, let $\vec{F}_D$ and $\vec{F}_L$ denote the drag and lift components of $\vec{F}_a$, i.e. 
\[\vec{F}_a := \vec{F}_L + \vec{F}_D,\]
with, by definition, $\vec{F}_L$ orthogonal to $\vec{v}_a$ and $\vec{F}_D$ parallel to $\vec{v}_a$. Consider also a (any) pair of angles $(\alpha,\beta)$ characterizing the orientation of $\vec{v}_a$ with respect to the body frame. The \emph{Buckingham $\pi-$theorem}~\cite[p. 34]{2010_AND} asserts that the intensity of the \emph{static} aerodynamic force varies like the square of the air speed $|\vec{v}_a|$  multiplied by a dimensionless function $C(\cdot)$ depending on the \emph{Reynolds number}
$R_e$, the \emph{Mach number} \emph{M}, and $(\alpha,\beta)$, i.e.
\begin{IEEEeqnarray}{RLL}
\label{buckingham}
k_a &:=& \frac{\rho \Sigma}{2}, \quad|\vec{F}_a|  =  k_a |\vec{v}_a|^2C(R_e,M,\alpha,\beta), 
\end{IEEEeqnarray}
with $\rho$ the \emph{free stream} air density, and $\Sigma$ an area germane to the given body shape. Then, further assuming 
that the direction of $\vec{F}_a$ depends upon the airspeed magnitude $|\vec{v}_a|$ via the $C(\cdot)$ function variables $(R_e,M)$ only and 
that this force does not (or little) depend(s) upon the angular velocity $\vec{\omega}$, one shows that this theorem in turn implies the existence of two dimensionless functions $C_D(\cdot)$ and $C_L(\cdot)$, and of a unit vector-valued function $\vec{r}(\cdot)$ characterizing the direction of the lift force w.r.t the body frame, such that
\begin{IEEEeqnarray}{L}
	\label{eq:FaComponents}
	\vec{F}_L = k_a|\vec{v}_a|C_L(R_e,M,\alpha,\beta)\vec{r}(\alpha,\beta) \times \vec{v}_a, \IEEEyessubnumber \\
	\vec{F}_D = -k_a |\vec{v}_a|C_D(R_e,M,\alpha,\beta)\vec{v}_a, \IEEEyessubnumber \\
        \vec{r}(\alpha,\beta) \cdot \vec{v}_a  =  0,  \IEEEyessubnumber \label{contR}
\end{IEEEeqnarray}
In the specialized literature $C_D$ ($\in \mathbb{R}^+$) and  $C_L$ ($\in \mathbb{R}$) are called the \emph{aerodynamic characteristics} of the body, and also the \emph{drag coefficient} and \emph{lift coefficient} respectively. 

\subsection{Aerodynamic models for symmetric bodies}

\begin{figure}[tb]
\centering
\def\svgwidth{.75\linewidth}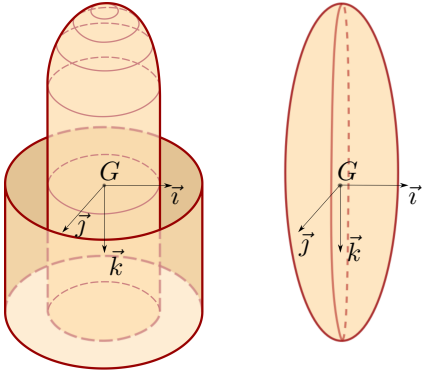
\caption{Symmetric and bisymmetric body shapes.}\label{fig:symmetric}
\end{figure}

The expressions \eqref{eq:FaComponents} of the lift and drag forces hold independently of the body's shape, since they are derived without any assumption upon this shape. However, 
shape symmetries of aerial vehicles --as well as of marine and ground vehicles-- are not coincidental. Simplification and cost reduction of the manufacturing process, despite their importance, are clearly not the main incentives accounting for the ubiquitous use of symmetric shapes. In this respect, Nature was first to give the example with most of the animals populating the Earth. On the basis of this observation, one could figure out numerous practical advantages resulting from symmetry properties. For flying purposes, however, not all symmetries are equally interesting. For instance, the sphere which represents the simplest most perfect symmetric 3D-shape is not best suited for energy-efficient long-distance flights because it does not allow for the creation of lift forces which can counteract the effects of gravity, in the same way --and almost as well-- as wheel-ground contact reaction forces for terrestrial vehicles, and buoyancy for marine and underwater vehicles. 
We here consider other kinds of symmetries in order to point out aerodynamic properties induced by them and their practical interest. More specifically, we focus on vehicles whose external surface $\mathcal{S}$ is characterized by the existence of
an orthonormal body frame $\mathcal{B} = \{G;\vec{\imath},\vec{\jmath},\vec{k}\}$, with $\vec{k}$ denoting the thrust direction according to Assumption \ref{hy:actuation}, that satisfies either one of the following assumptions. 
\begin{hypothesis}[\textbf{Symmetry}]
  \label{hy:symmetries}
  Any point $P \in \mathcal{S}$ transformed by the rotation of an angle $\theta$ about the axis $G\vec{k}$, i.e. by the operator defined by
  \[ g_\theta(\cdot) = \rot_{G\vec{k}}(\theta)(\cdot),\]
  also belongs to $\mathcal{S}$, i.e. $g_\theta(P) \in \mathcal{S}$.
\end{hypothesis}
\begin{hypothesis}[\textbf{Bisymmetry}]
  \label{hy:bisymmetries}
  Any point $P \in \mathcal{S}$ transformed by the composition of two rotations of angles $\theta$ and $\pi$ about the axes $G\vec{k}$ and $G\vec{j}$, i.e. by the operator defined by
  \[ g_\theta(\cdot) = (\rot_{G\vec{k}}(\theta) \circ \rot_{G\vec{\jmath}}(\pi))(\cdot),\]
  also belongs to $\mathcal{S}$, i.e. $g_\theta(P) \in \mathcal{S}$.
\end{hypothesis}
The operator $\rot_{O\vec{v}}(\xi)(P)$ stands for the rotation about the axis $O\vec{v}$ by the angle $\xi$ of the point $P$.
Examples of ``symmetric'' and ``bisymmetric'' shapes satisfying these assumptions are represented in Figure~\ref{fig:symmetric}.

\begin{figure}[tb]
\centering
\vspace*{15mm}
\def\svgwidth{.78\linewidth}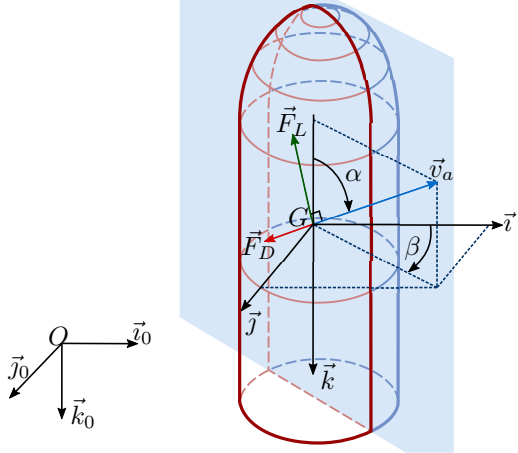
\caption{Aerodynamic forces and $(\alpha,\beta)$ angles.}
\label{fig:aerodyEff}
\end{figure}

For symmetric shapes, i.e. such that Assumption~\ref{hy:symmetries} holds true, one can define $\alpha \in [0,\pi]$ as the {\em angle of attack}\footnote{The angle of attack $\alpha$ so defined does not coincide with the one used for airplanes equipped with flat wings which break the body's rotational symmetry about $G\vec{k}$~\cite[p. 53]{2004_STENGEL}.} between $-\vec{k}$ and $\vec{v}_a$, and $\beta \in (-\pi,\pi]$ as the angle between the unit frame vector $\vec{\imath}$ and the projection of $\vec{v}_a$ on the plane $\{G;\vec{\imath},\vec{\jmath}\}$ (see Fig. \ref{fig:aerodyEff}). Observe that this assumption also implies that:

$\mathbf{P1:}$ the aerodynamic force $\vec{F}_a$ does not change when the body rotates about its axis of symmetry $G\vec{k}$;
$\\$ \noindent
$\mathbf{P2:}$ $\vec{F}_a \in \text{span}\{\vec{k},\vec{v}_a\}$ .

Property P1 in turn implies that the aerodynamic characteristics do not depend on $\beta$, whereas Property P2 implies that $i)$ $\vec{r} \ $ is orthogonal to $\vec{k}$, $ii)$ $\vec{r}$ is independent of $\alpha$, and $iii)$ the lift coefficient is equal to zero when $\alpha = \{0,\pi\}$. 
Subsequently, the expressions~\eqref{eq:FaComponents} of the lift and drag forces specialize to
\begin{IEEEeqnarray}{L}
	\label{eq:aerodyModelSymmetricBodies}
	\vec{F}_L = k_a|\vec{v}_a|C_L(R_e,M,\alpha)\vec{r}(\beta) \times \vec{v}_a, \IEEEyessubnumber \\
	\vec{F}_D = -k_a |\vec{v}_a|C_D(R_e,M,\alpha)\vec{v}_a, \IEEEyessubnumber \\
        \vec{r}(\beta)=-\sin(\beta) \vec{\imath}+\cos(\beta) \vec{\jmath}. \IEEEyessubnumber
\end{IEEEeqnarray}
Under the stronger Assumption~\ref{hy:bisymmetries}, i.e. when the body's shape is also $\pi$-symmetric w.r.t. the $G\vec{\jmath}$ axis, the aerodynamic characteristics $C_L$ and $C_D$ must be $\pi-$periodic w.r.t. $\alpha$.
The aforementioned choice of $(\alpha,\beta)$ implies that 
\begin{IEEEeqnarray}{c}
	\label{system:alphaBeta}
	\alpha = \cos^{-1}\left(-\frac{v_{a_3}}{|\vec{v}_a|}\right),\quad 
	\beta = \mathrm{atan2}(v_{a_2},v_{a_1}),
	\label{eq:Beta}
\end{IEEEeqnarray}
and
\begin{equation}
	\label{system:vaDecomposition}
	\begin{system}
		v_{a_1} &=& |\vec{v}_a|\sin(\alpha)\cos(\beta), \\
		v_{a_2} &=& |\vec{v}_a|\sin(\alpha)\sin(\beta), \\
		v_{a_3} &=& -|\vec{v}_a|\cos(\alpha).
	\end{system}
\end{equation}
From the definitions of $\alpha$ and $\vec{r}(\beta)$, one then verifies that
\[\vec{r}(\beta) \times \vec{v}_a = -\cot(\alpha) \vec{v}_a-\frac{|\vec{v}_a|}{\sin(\alpha)}\vec{k},
\]
so that $\vec{F}_a = \vec{F}_L+\vec{F}_D$ becomes
\begin{IEEEeqnarray}{RCL}
\label{FaNewExpressSymm}
\vec{F}_a &=& {-}k_a |\vec{v}_a|\hspace{-0.1cm}\left[\hspace{-0.1cm} \Big(\hspace{-0.05cm}C_D(\cdot)+C_L(\cdot)\cot(\alpha)\Big)\vec{v}_a+\frac{C_L(\cdot)}{\sin(\alpha)}|\vec{v}_a| \vec{k} \right]. \IEEEeqnarraynumspace \nonumber
\end{IEEEeqnarray}
\subsection{Aerodynamic models yielding spherical equivalency}
For constant Reynolds and Mach numbers the aerodynamic coefficients depend only on 
$\alpha$.
By using the above relationship, it is a simple matter to establish the following result.

\begin{proposition}
\label{th:conditionA}
Consider a symmetric thrust-propelled vehicle. Assume that
the aerodynamic forces are given by \eqref{eq:aerodyModelSymmetricBodies} and that the aerodynamic coefficients satisfy the following relationship
\begin{IEEEeqnarray}{r}
		\label{eq:conditionOnalpha}
		C_D(\alpha)+ C_L(\alpha)\cot(\alpha) = C_{D_0},  \IEEEeqnarraynumspace
\end{IEEEeqnarray}
with $C_{D_0}$ denoting a constant number.
Then the body's dynamic equation \eqref{eq:newton} may also be written as
\begin{IEEEeqnarray}{c}
m\vec{a}=m\vec{g}+ \vec{F}_p -T_p\vec{k}, 
\label{eq:newFormDynamics} 
\end{IEEEeqnarray}
with
\begin{IEEEeqnarray}{RCL}
T_p &=& T + k_a|\vec{v}_a|^2\frac{C_L(\alpha)}{\sin(\alpha)},
\label{eq:TpTh} \IEEEyessubnumber \IEEEeqnarraynumspace \\ 
\vec{F}_p &=& -k_a C_{D_0}|\vec{v}_a| \vec{v}_a. \IEEEyessubnumber  \IEEEeqnarraynumspace
\label{eq:fpGen}
\end{IEEEeqnarray}
\end{proposition}
The important result is the non-dependence of $\vec{F}_p$ upon the angle of attack $\alpha$, and thus on the vehicle's orientation.
The interest of this proposition is to point out the possibility of seeing a symmetric body subjected to both drag and lift forces as a sphere subjected to an equivalent drag force $\vec{F}_p$ and powered by an equivalent thrust force $\vec{T}_p=-T_p\vec{k}$. 
The main condition is that the relation \eqref{eq:conditionOnalpha} must be satisfied. 
Obviously, this condition is compatible with an infinite number of functions $C_D$ and $C_L$. 
Let us 
point out  a particular set of simple functions, already considered in the 2D-case addressed in \cite{2011_pucci}, which also satisfy the $\pi$-periodicity property w.r.t. the angle of attack $\alpha$ associated with bisymmetric bodies. 

\begin{proposition}
\label{prop:2D->3D}
The functions $C_D$ and $C_L$ defined by 
\begin{equation}
		\label{system:aerodynamicsChaPart}
		\begin{system}
            C_D(\alpha) &=& c_0 + 2c_1\sin^2(\alpha) \\
			C_L(\alpha) &=&  c_1\sin(2\alpha)
		\end{system}
\end{equation}
with $c_0$ and $c_1$ 
two real numbers, satisfy the condition \eqref{eq:conditionOnalpha} with 
$C_{D_0} = c_0 + 2c_1.$ 
The equivalent drag force and thrust intensity are then given by
\begin{IEEEeqnarray}{RCL}
		\label{eq:parapetersNFD}
		\vec F_p &=& -k_a(c_0 + 2c_1)|\vec{v}_a| \vec{v}_a, \IEEEyessubnumber \\ T_p &=& T + 2c_1k_a|\vec{v}_a|^2\cos(\alpha). \IEEEyessubnumber \IEEEeqnarraynumspace
\end{IEEEeqnarray}
\end{proposition}

\begin{figure}[t]
 \centering
 \small{\input{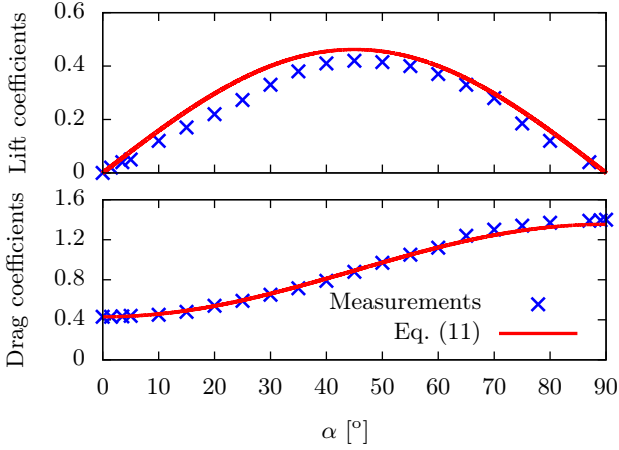}}\vspace*{-1.25cm}
 \caption{Aerodynamic coefficients of elliptic shaped bodies.}
 \label{fig:ellipticShaped}
\end{figure}

\begin{figure}[t]
 \centering
 \small{\input{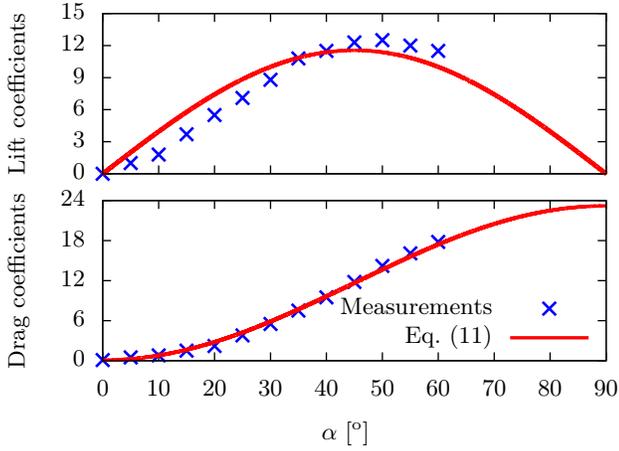}}\vspace*{-1.25cm}
 \caption{Aerodynamic coefficients of missile-like bodies.}
 \label{fig:missileCharacteristics}
\end{figure}

A particular bisymmetric body is the sphere whose aerodynamic characteristics (zero lift and constant drag coefficient) are obtained by setting $c_1=0$ in \eqref{system:aerodynamicsChaPart}.
Elliptic-shaped bodies are also symmetric but, by contrast with the sphere, they do generate lift in addition to drag. The process of approximating measured aerodynamic characteristics with functions given by \eqref{system:aerodynamicsChaPart} is illustrated by the Figure~\ref{fig:ellipticShaped} where we have used experimental data borrowed from \cite[p.19]{1965_WAYNE} for an elliptic-shaped body with Mach and Reynolds numbers equal to $M = 6$ and $R_e = 7.96\cdot 10^6$ respectively. For this example, the identified coefficients are $c_0=0.43$ and $c_1=0.462$. Since missile-like devices are ``almost'' bisymmetric, approximating their aerodynamic coefficients with such functions can also be attempted. For instance, the approximation shown in Figure~\ref{fig:missileCharacteristics} has been obtained by using experimental data taken from \cite[p.54]{1971_SAFFEL} for a missile moving at $M = 0.7$. In this case, the identified coefficients are $c_0=0.1$ and $c_1=11.55$. In both cases, the match between 
experimental data and the approximating functions, although not perfect, should be sufficient for feedback control purposes.

Other functions satisfying the condition \eqref{eq:conditionOnalpha} may also be of interest to model the aerodynamic characteristics of other, almost bisymmetric, bodies. For instance, by considering annular wings, functions of the form \[C_D(\alpha)=\bar{c}_0, \quad C_L(\alpha)=\bar{c}_1 \tan(\alpha),\] 
with $\bar{c}_0$ and $\bar{c}_1$ denoting two real numbers, can be used to model classical linearly increasing lift and quasi-constant drag at small angles of attack, i.e. before the stall region is reached, in combination with functions \eqref{system:aerodynamicsChaPart} that are more representative of the physics for angles beyond the stall region. The possibility of forming in this way more complex functions that apply to variously shaped bodies and that (almost) satisfy the condition \eqref{eq:conditionOnalpha} over a large range of angles of attack is the subject of complementary studies.     

\section{Control}
\label{sec:controlDesign}

To illustrate the interest of the transformation evoked previously, let us consider the problem of stabilizing a reference velocity $\vec{v}_r=(\vec{\imath}_0,\vec{\jmath}_0,\vec{k}_0)\dot{x}_r=(\vec{\imath},\vec{\jmath},\vec{k})v_r$ asymptotically. The application of the control solution proposed in \cite[Sec. III.D]{2009_HUA} to System \eqref{eq:newton}-\eqref{eq:newtonM}, with $(\vec{F}_a,T)$ replaced by the equivalent drag force and thrust intensity $(\vec{F}_p,T_p)$ defined in Proposition \ref{th:conditionA}, yields the following control expressions
\begin{IEEEeqnarray}{RCL}
		\label{controlV}
	 	T & = & \bar{f}_{a_3} + k_1 |f_p|\tilde{v}_3, \IEEEyessubnumber \IEEEeqnarraynumspace \\
		\omega_1 & = & -k_2 |f_p| \tilde{v}_2-\frac{k_3|f_p| \bar{f}_{p_2}}{(|f_p|+\bar{f}_{p_3})^2}+\frac{\bar{f}_p^TS(e_1)R^T\dot{f}_p}{|f_p|^2}, \IEEEyessubnumber \IEEEeqnarraynumspace \\
		\omega_2 & = & k_2 |f_p|\tilde{v}_1+\frac{k_3|f_p| \bar{f}_{p_1}}{(|f_p|+\bar{f}_{p_3})^2}-\frac{\bar{f}_p^TS(e_2)R^T\dot{f}_p}{|f_p|^2}, \IEEEyessubnumber \IEEEeqnarraynumspace
\end{IEEEeqnarray}
with $\tilde{v}:= v-v_r$, $\vec{a}_r:=\tfrac{d}{dt}\vec{v}_r=(\vec{\imath}_0,\vec{\jmath}_0,\vec{k}_0)\ddot{x}_r$,
\begin{IEEEeqnarray}{RCLRCLRCL}
  \label{eq:fAndfp}
  \vec{f}_a &=&  (\vec{\imath}_0,\vec{\jmath}_0,\vec{k}_0)f_a &=&  (\vec{\imath},\vec{\jmath},\vec{k})\bar{f}_a  &:=& m\vec{g}+\vec{F}_a-m \vec{a}_r,
  \IEEEyessubnumber \IEEEeqnarraynumspace \\
  \vec{f}_p &=&  (\vec{\imath}_0,\vec{\jmath}_0,\vec{k}_0)f_p &=&  (\vec{\imath},\vec{\jmath},\vec{k})\bar{f}_p  &:=& m\vec{g}+\vec{F}_p-m \vec{a}_r,  \IEEEyessubnumber \IEEEeqnarraynumspace
\end{IEEEeqnarray}
and $k_{1,2,3}$ three positive real numbers.
Note that, using \eqref{eq:fpGen}, the vector $f_p$ of coordinates of $\vec{f}_p$ expressed in the fixed frame ${\mathcal I}$ is equal to $mge_3-k_aC_{D_0}|v_a|\dot{x}_a-m\ddot{x}_r$, and is thus independent of the vehicle's orientation. Therefore, its time-derivative does not depend on the angular velocity vector $\omega$ and the above expressions of the first two components of this vector are well defined. The interest of the invoked transformation, combined with \eqref{eq:conditionOnalpha}, lies precisely there.
As for the last component $\omega_3$, since it does not influence the vehicle's longitudinal motion due to the symmetry about the axis $G\vec{k}$, it does not have to be defined 
 at this point. This free degree of freedom can be used for complementary purposes involving, for instance, the  angle $\beta$.

Let $\tilde{\theta} \in (-\pi,\pi]$ denote the angle between $\vec{k}$ and $\vec{f}_p$. In \cite{2009_HUA}, stability and convergence properties associated  with the feedback control \eqref{controlV} are established by using the Lyapunov function candidate
\[V = \frac{|\tilde{v}|^2}{2} + \frac{1}{k_2m} \big[ 1-\cos(\tilde{\theta}) \big],\] whose time-derivative along any trajectory of the controlled system is
\[\dot{V} = -k_1|f_p|\tilde{v}^2_3 - \frac{k_3}{k_2} \tan^2\left(\tilde{\theta}/2\right).\]
Assuming that $\vec{v}_w$ and $\vec{v}_r$ are bounded in norm up to their second time-derivatives, and provided that $\exists\,\delta>0$ such that $|f_p|>\delta$, $\forall t \in \mathbb{R}^+$, one shows that the equilibrium $(\tilde{v}, \tilde{\theta})=(0,0)$ of the controlled system is asymptotically stable, with the domain of attraction equal to $\mathbb{R}^3  \times (-\pi,\pi)$.

In practice, the control law must be complemented with integral correction terms to compensate for almost constant unmodeled additive perturbations.
The solution proposed in \cite{2009_HUA} involves $\vec{I}_v =  (\vec{\imath}_0,\vec{\jmath}_0,\vec{k}_0)I_v$ with
$I_{v} :=\int_0^{t}\dot{\tilde{x}}(s)\,ds$,
and $\dot{\tilde{x}} := R \tilde{v}$ the longitudinal velocity error expressed in the inertial frame. Also, let $h$ denote a smooth bounded strictly positive function defined
on $[0,+\infty)$ satisfying the following properties (\cite[Sec. III.C]{2009_HUA}) for some positive constant numbers $\eta,\mu$,
$\forall s \in \mathbb{R}, \quad | h(s^2)s | <\eta \; \text{and} \; 0<\frac{\partial }{\partial s}(h(s^2)s) <\mu$.
It then suffices to replace the definitions \eqref{eq:fAndfp} of $\vec{f}_a$ and $\vec{f}_p$ by 
\begin{IEEEeqnarray}{RCL}
  \label{eq:fAndfpWithIt}
  \vec{f}_a &:=& m\vec{g}+\vec{F}_a-m \vec{a}_r +  h(|{I}_v|^2)\vec{I}_v
  \IEEEyessubnumber \IEEEeqnarraynumspace \\
  \vec{f}_p &:=& m\vec{g}+\vec{F}_p-m \vec{a}_r+  h(|{I}_v|^2)\vec{I}_v \IEEEyessubnumber  \label{fpWithIt} \IEEEeqnarraynumspace
\end{IEEEeqnarray}
in \eqref{controlV} to obtain a control which incorporates an integral correction action and for which strong stability and convergence properties can also be proven (\textit{c.f.} \cite{2009_HUA}).

\begin{remark}
  As in the case of velocity control, the position control presented in \cite{2009_HUA} can be adapted to stabilize reference trajectories. Videos of simulations with several reference trajectories can be found at {\tt http://goo.gl/HKQtz}
\end{remark}

\section{Conclusion and perspectives} \label{conclusion}
The paper sets basic principles for the modeling and nonlinear control of aerial vehicles subjected to strong aerodynamic forces. Possible extensions are numerous. 
They concern in particular airplanes and other vehicles whose lift properties mostly rely on the use of large flat surfaces (wings) which break the body symmetries here considered. The dreaded stall phenomenon, when it is pronounced, is not compatible with transformations alike these evoked in the paper, because it forbids the existence of equivalent drag forces that do not depend on the vehicle's angle of attack. Nor is it even compatible with the uniqueness of cruising equilibria and the objective of asymptotic stabilization, as pointed out in \cite{2012_PUCCI}. Its importance at both the modeling and control levels, and its consequences during transitions between hovering and lift-based-cruising, need to be studied and, if possible, attenuated via an adequate control design. Clearly, the control solution here proposed also calls for a multitude of complementary extensions and adaptations before it is implemented on a physical device.  Let us just mention the production of the control torque allowing for 
desired angular velocity changes and the determination of corresponding low level control loops that take actuators' physical limitations into account --in relation, for instance, to the airspeed dependent control authority associated with the use of flaps and rudders. Measurement and estimation of various physical variables involved in the calculation of the control law, other than the ever needed information about the vehicle's position and attitude, such as the air velocity and the angle of attack, or the thrust force produced by a propeller, also involves a combination of hardware and software issues which are instrumental to implementation.     

\bibliographystyle{plain}        
\bibliography{bibliography}           

\end{document}